\documentclass[journal]{IEEEtran}
\usepackage[table]{xcolor}
\usepackage{graphicx,times,amsmath,cite,enumerate}
\usepackage{multirow}
\usepackage{tabularx}
\usepackage{kantlipsum}
\usepackage{threeparttable}
\usepackage{epsfig}
\usepackage{amssymb}
\usepackage{enumitem}
\usepackage{latexsym}
\usepackage{psfrag}
\usepackage{setspace}
\usepackage{color}
\usepackage{epstopdf}
\usepackage{verbatim}
\usepackage{subfigure}
\usepackage{footnote}
\usepackage{array}
\newcolumntype{P}[1]{>{\centering\arraybackslash}p{#1}}
\usepackage{float}
  
\usepackage{multirow}
\usepackage{textcomp}
\PassOptionsToPackage{hyphens}{url}
\usepackage{url}
\usepackage{physics}
\usepackage{hyperref}
\usepackage{etoolbox}
\usepackage[ruled,vlined]{algorithm2e}
\let\oldemptyset\emptyset
\let\emptyset\varnothing
\usepackage{nomencl}
\usepackage[edges]{forest}
\usetikzlibrary{arrows.meta}
\usepackage{url}
\begin{document}

\title{Causative Cyberattacks on Online Learning-based Automated Demand Response Systems}
\author{Samrat Acharya, \emph{Student Member, IEEE}, Yury Dvorkin, \emph{Member, IEEE},
        and~Ramesh Karri, \emph{Fellow, IEEE}}

\maketitle
 \begin{abstract} Power utilities are adopting Automated Demand Response (ADR) to replace the costly fuel-fired generators and to preempt congestion during peak electricity demand. Similarly, third-party Demand Response (DR) aggregators are leveraging controllable small-scale electrical loads to provide  on-demand grid support services to the utilities. Some aggregators and utilities have started employing  Artificial Intelligence (AI) to learn the energy usage patterns of electricity consumers and use this knowledge to design optimal  DR incentives. Such AI frameworks use open communication channels between the utility/aggregator and the DR customers, which are vulnerable  to \textit{causative} data integrity cyberattacks. This paper explores vulnerabilities of AI-based DR learning and designs a data-driven attack strategy informed by DR data collected from the New York University (NYU) campus buildings. The case study demonstrates the feasibility and effects of maliciously tampering with (i)  real-time DR incentives, (ii)  DR event data sent to DR customers, and (iii)  responses of DR customers to the DR incentives.  
 \end{abstract}
 
\IEEEpeerreviewmaketitle

\section{Introduction}
\label{intro}

\IEEEPARstart{I}{nternet-of-Things} (IoT) enable power utilities to adopt Demand Response (DR) strategies to reduce or shift electricity demand peaks. Thus, in 2018,  the US  utilities used $\sim$4.5\% of the peak load capacity for DR services. This is estimated to increase to 20\% in 2030 yielding annual cost savings of $>$\$15 billion \cite{sepa}. The DR resources vary from electricity-dependent industrial processes (e.g., steel industry and irrigation control) to  high-wattage  residential  appliances (e.g.,  Electric Vehicles (EVs), air-conditioners, and refrigerators). DR is either managed directly by the  utilities or by third-party, for-profit DR aggregators. For example, there are  $20+$ third-party aggregators in California, US that cooperate with local power utilities \cite{DR_caiso, PGNE}.

The utilities and aggregators are automating interactions with their DR-enabled customers. By default, Automated Demand Response (ADR) does not require human intervention to optimize, schedule, and exercise DR capacity. However, interventions are possible. The most common standard for  automated communication between DR customers and an aggregator or utility is the Open Automated Demand Response (OpenADR), which is recognized by the  International Electrotechnical Commission (IEC)  \cite{open_adr}. The OpenADR  standard defines the two ends of a communication channel as the Virtual Top Node (VTN) and the Virtual End Node (VEN), issues digital certificates to the nodes for an authenticated communication, and encrypts information exchanged between the nodes. Despite the standardization and the use of industry-grade encryption, cyber threats in ADR prevail because DR customers lack industry-grade cyber defense and hygiene on their devices.

Widely deployed Smart Meters (SMs) and  Building Energy Management Systems (BEMSs), i.e., VENs in the OpenADR framework are vulnerable to cyberattacks. SMs and BEMSs use wireless communication technologies (e.g., WiFi, ZigBee, BACnet, and ModBus) to interface within a co-located area or a building and use Wide Area Networks (WAN) such as cellular networks and Power Line Communication (PLC) to interface with the utilities or aggregators \cite{qi2017demand, mahmood2016lightweight, alphagranular}. Attackers can exploit vulnerabilities in the communication technologies 
 (e.g., WiFi \cite{wifi}, ZigBee \cite{fan2017security}) to  compromise SMs and BEMSs. On the other hand, cellular networks and PLC are vulnerable to man-in-the-middle cyberattacks \cite{seijo2017cybersecurity}.
 
In addition to vulnerabilities in the ADR, SMs are vulnerable to remote cyberattacks compromising data privacy, causing Denial-of-Service (DoS), and enabling power theft \cite{weaver2017smart, smartmeter,kumar2019security, tabrizi2019design}. Physical accessibility of SMs increases their attack surface, which can be defined as the total number of vulnerable devices and processes. Tabrizi et al. \cite{tabrizi2019design} acknowledge the importance of physical access to SMs for successful DoS and data tampering attacks.  Attack surfaces of SMs and BEMSs can be increased by malicious smart appliances (e.g., smart television, smart lights and EVs \cite{acharya2020cybersecurity}) connected to the same network. Since these smart appliances have complex, opaque and porous  supply chains, attackers can exploit these supply chains to compromise SMs and BEMSs. For instance, the US Federal Bureau of Investigation (FBI) alarmed the private sector in February 2020 about Kwampirs, a remote access malware that infects the supply chains of the software products related to the US energy sector  \cite{fbi_alert}. Although these attacks have not been encountered at scale, their possibility has lead to investigating the feasibility of demand-side cyberattacks on power grids. 

 Recent studies \cite{acharyaplugin,soltan2018blackiot,pasqualetti2019} considered demand-side cyberattacks on power grids  launched by manipulating EVs and Heating, Ventilation, and Air-Conditioning (HVAC) units. Acharya et al. \cite{acharyaplugin} demonstrated that public power grid and commercial EV Charging Stations (EVCSs) data can be weaponized to launch remote attacks on the power grid. The study collected public data from documents and websites of the local power utility and their affiliates in Manhattan, New York. The main result in  \cite{acharyaplugin} is that a botnet compromising $\approx 1000$ modern EVs charging at $350$ kW EVCSs can  cause over-frequency in the  power grid, tripping a  generator cascading into an urban brownout. Authors in \cite{soltan2018blackiot,pasqualetti2019} analyzed demand-side cyberattacks launched by tampering with generic HVAC appliances. These studies were performed on IEEE test systems and reported frequency instability  and line overloading, causing cascading outages and a dramatic increase in the operating costs for the power grid. Raman et al. \cite{raman} presented attacks manipulating behavior of DR customers, with the intention of depleting system reserves and corrupting voltage profiles. The common thread of the attacks in \cite{acharyaplugin,soltan2018blackiot,pasqualetti2019,raman} is their focus on manipulations of demand-side appliances during DR events assuming that an utility/aggregator  does not use AI  to learn and monitor DR behavior. With the proliferation of DR customers and advances in AI, such learning and monitoring solutions have become available  \cite{khezeli2017risk, li2017distributed,robert, 9125942} and practiced (e.g., \cite{enpower}). 

  Khezeli and Bitar \cite{khezeli2017risk} have developed a learning scheme to dynamically calculate the monetary incentives paid to DR customers  to minimize the risk exposure of the utility facing an a priori uncertain response of DR customers. Li et al. \cite{li2017distributed} extended the approach in \cite{khezeli2017risk} to a distributed learning environment using the  aggregated DR response of customers, thus avoiding the need to observe  individual DR customers. In Meith and Dvorkin \cite{robert} and Tucker et al. \cite{9125942} the approaches in \cite{khezeli2017risk} and \cite{li2017distributed} have been extended to accommodate learning subject to distribution network  constraints. 
The common thread in \cite{khezeli2017risk,li2017distributed,robert,9125942} is that learning is myopic to cyber vulnerabilities of the DR resources, thus assuming that monetary incentives generated by learning are benign, i.e.,  DR operations are not affected by corrupt data. In contrast, this paper investigates cyberattacks on the DR learning\footnote{Information required for the aggregated learning is publicly available. The aggregated DR data and incentives are available via smartphone apps to the DR customers. Public data is an attack vector in the urban power grids, \cite{acharyaplugin}.} schemes (e.g., in \cite{khezeli2017risk,li2017distributed,robert,9125942}), which are known as \textit{causative attacks} in the  AI and cybersecurity communities.

Causative attacks denote malicious alterations of input or training data that cause a malfunction of the underlying learner  and lead to unintended, erroneous  learning outcomes \cite{mei2015using, liu2017iterative}. 
 Mei and Zhu \cite{mei2015using} presented a framework to identify possible causative attacks on the learner via a bi-level optimization model, where the upper level models a learner and the lower level models an attacker.  The bi-level optimization is solved offline to  return the optimal training data alteration to achieve a given attack goal, which is then  fed into a targeted decision-making process.  Liu et al. \cite{liu2017iterative} consider online causative attacks on machine teaching, where an attacker sequentially feeds the malicious training data to the learner. Online causative attacks are more relevant for the DR learning context because DR operations are sequential in nature and benefit from an iterative learning process (e.g., as  presented in \cite{khezeli2017risk,li2017distributed,robert}).
 
In the context of related cyberattacks summarized in Table~\ref{tab:related_attack},  this paper makes  three  contributions:
\begin{enumerate}
\item It analyzes cyber vulnerabilities of a state-of-the-art DR learning set up that can be used by either a power utility or an aggregator.
\item It studies impacts of specific attributes of DR programs such as DR events and DR customers on  attack performance using the Shapley value theorem \cite{shapley1953value,castro2009polynomial,jia2019towards,ghorbani2019data}, which  gives an equitable, fair, and rational estimate of the attack sensitivity to each DR attribute.
\item Using real-life data on DR power curtailment and incentives from the NYU microgrid and buildings, it simulates an attack under conditions similar to  US electric power distribution systems to assess the impact of causative attacks on the monetary incentives designed by a utility/DR aggregator. Furthermore, technical challenges such as changes in peak demand and frequency excursions imposed to the power utility are investigated. Finally, we anticipate that the paper raises   awareness about implications of causative cyberattacks on ML-enabled and IoT-powered DR programs, thus contributing to striking the consensus among  concerned parties (e.g., regulators, power utilities, third-party aggregators, and electricity consumers). 
\end{enumerate}
  \begin{table}[!t]
 \centering
 \centering
 \caption{Related Demand-side (DS) and Causative (C) Cyberattacks}
 \begin{threeparttable}[!t]
	\centering
	\resizebox{0.98\columnwidth}{!}{
	
    \begin{tabular}{|p{2.2cm}| p{0.25cm} p{0.28cm}| p{5.9cm}|}
    
    \hline
    &\multicolumn{2}{c|}{\textbf{Attack}}&\textbf{\centering Attack Vector} \\ 
    
    \multirow{1}{*}{\textbf{Paper}} &{\rotatebox{90}{DS }}&{\rotatebox{90}{C}}&\\
    \hline
    \rule{0pt}{2ex}\cite{fan2017security,seijo2017cybersecurity}& \checkmark& & Communication protocols, e.g., WiFi, ZigBee.\\ \hline
        \rule{0pt}{2ex}\cite{weaver2017smart, smartmeter,kumar2019security, tabrizi2019design}& \checkmark& & Smart Meters (SMs) and IoT devices. \\ \hline
        \rule{0pt}{2ex}\cite{acharyaplugin}& \checkmark& & Public data on EVs and power grid.\\ \hline
        \rule{0pt}{2ex}\cite{soltan2018blackiot,pasqualetti2019}& \checkmark& & Generic demand-side high-power devices.\\ \hline
         \rule{0pt}{2ex}\cite{raman}&& \checkmark& Behavior of DR customers.\\ \hline
          \rule{0pt}{2ex}\cite{mei2015using}&& \checkmark& Training data for underlying offline Machine Learning (ML).\\ \hline
            \rule{0pt}{2ex}\cite{liu2017iterative}&& \checkmark& Training data for underlying online ML.\\ \hline
                       \rule{0pt}{2ex}This paper&\checkmark& \checkmark& Causative demand-side attack on  DR ML.\\
       \hline
    \end{tabular}
    }
    \end{threeparttable}
    \label{tab:related_attack}
 \end{table}

\section{Demand Response Framework in Smart Grids}
\subsection{DR Architecture}
\label{sec:DR_framework}
\begin{figure}[t!]
\centering
\includegraphics[width=\columnwidth, clip=true, trim= 0mm 0mm 0mm 0mm]{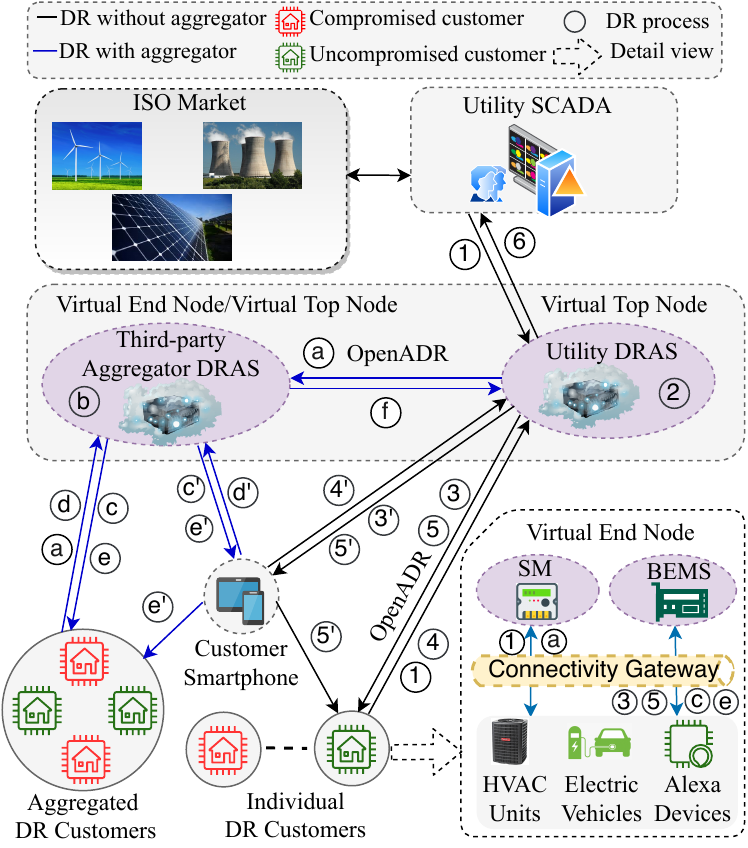}
    \caption{ML-enabled residential DR program with processes (PR.).}
    \label{fig:DR_communication}
\end{figure}

 Fig.~\ref{fig:DR_communication} shows a schematic of the learning-based ADR program employed by either an utility or a third-party DR aggregator. The DR program has six processes (PR.). First, a Demand Response Automated Server (DRAS)  acquires real-time energy usage data of DR customers from their individual SMs via the cellular networks and PLC (PR. 1).  DRAS receives power grid operational schedules from the utility or, in rare instances, from the Independent System Operator (ISO) via the fiber optic communication lines and cellular networks (PR. 1). Second, the DRAS runs a DR scheduling and pricing algorithm (e.g., heuristic or ML-based) on the collected SM and power grid data (PR. 2). Third, the DR price signals and schedules are sent to the DR customers using  OpenADR 2.0 protocol\footnote{In OpenADR, the DRAS is a VTN and the DR customer is a VEN.}  (PR. 3). Alternatively, this information is sent to the DR customers via the customer smartphone app (PR. 3'). Fourth, the DRAS or the VTN receives the responses of the DR customers to the DR call via the OpenADR protocol (PR. 4). This response includes an acknowledgement of receipt of the DR call, accept/reject notice to participate in the  DR event, and an estimate of the power available from that VEN. Alternatively, the DR customers can respond via a smartphone app (PR. 4').  Fifth, after accepting the DR call, the VTN can control the DR resources registered in the ADR program, by directly communicating/coordinating with a local BEMS (PR. 5). DR customers can overrule the ADR commands generated by the VTN and control the real-time operations of DR resources  via a smartphone app (PR. 5'). Finally, the DRAS sends real-time DR information to the utility (PR. 6). 

As shown in Fig.~\ref{fig:DR_communication}, third-party aggregators use DR techniques (PR. a-f) similar to the utilities with two exceptions. First, the aggregator DRAS bids and reports available DR services to the utility DRAS, instead of the ISO market. Second, they use proprietary communication protocols that maybe vulnerable to cyberattacks. The DR aggregators are vital as power utilities have started commonly employing for-profit, third-party aggregators as DR providers to serve three purposes: i) to aggregate small-scale and distributed residential DR resources and provide flexibility to the utility at bulk, ii) to transfer the risk associated with DR uncertainty to the aggregators, and (iii) in some cases, to intensify  distribution deregulation  \cite{burger2017review, mieth2021learning}.

\subsection{Causative Attack on ML-Based DR}
\label{sec:attack_description}
\begin{figure}[!t]
\centering
\includegraphics[width=\columnwidth, clip=true, trim= 13mm 0mm 0mm 0mm]{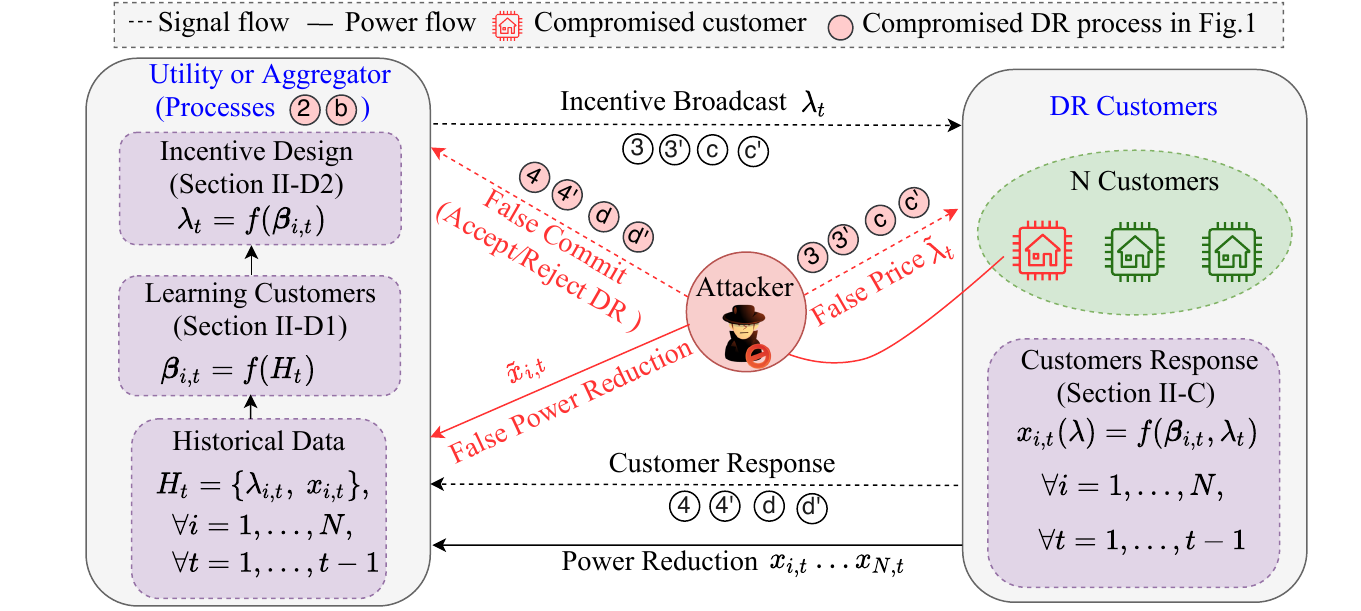}
    \caption{A causative attack on the learning-based DR schemes used by utilities and aggregators. The  malicious (or compromised) DR customers are in red and the benign customers are in green.}
    \label{fig:attack}
\end{figure}

Fig.~\ref{fig:attack} shows the vulnerabilities introduced by the participants in a DR program. These vulnerabilities can be used to launch causative attacks on ML-enabled DR in Fig.~\ref{fig:DR_communication}. For instance, an attacker can tamper with the DR schedules and incentives sent to the DR customers (PR. 3, 3', c, and c'). They can manipulate the response of the DR customers to the DR calls (PR. 4, 4', d, and d'). They can interfere with the power reduced by DR customers during the DR events. By feeding erroneous signals and data to the utility/aggregator DRAS ML model (PR. 2 and b), it impedes the model from learning the true behaviors of DR customers. The model may recommend subpar DR incentives undermining the DR program.

\subsection{DR Customer Model}
\label{sec:DR_customer_model}
This section provides a mathematical model of an individual  DR customer based on \cite{robert}. The utility function of a DR customer can be modeled as a quadratic function of its power consumption \cite{samadi2012advanced,robert,khezeli2017risk,li2017distributed}:
\begin{equation}
\label{eq:utility}
U_{i,t}(x, \alpha) =\frac{1}{2}\alpha_{1,i,t} x_{i,t}^2 +\alpha_{0,i,t}x_{i,t},
\end{equation}
where $x_{i,t}$ is the power consumed by a DR customer $i$ at time $t$, and $\alpha_{1,i,t}$, $\alpha_{0,i,t}$ are coefficients representing the energy use behavior of the $i^{th}$ DR customer at time $t$. This utility function  quantifies the \textit{satisfaction} of a DR customer $i$ for consuming power $x_{i,t}$  at time $t$. Alternatively, it is a measure of cost or \textit{dissatisfaction} of the DR customer $i$ for not consuming power $x_{i,t}$  at time $t$. As explained in Section~\ref{sec:DR_framework}, DR programs make it possible for  DR customers to define \textit{dissatisfaction} tolerance and incentivize them. 

In a typical DR program, an utility/aggregator broadcasts a DR incentive $\lambda_t \geq 0$ (\$/kWh) that the DR customer receives by taking part in a DR event scheduled for time $t$. By agreeing to reduce its power consumption by $x_{i,t}$  (kW), the DR customer $i$ receives a reward $\lambda_tx_{i,t}$ (\$). Each DR customer optimizes its trade-off between the reward received and the \textit{dissatisfaction} incurred by power reduction. The objective function of DR customer $i$ for a DR event at time $t$ is:
\allowdisplaybreaks
\begin{equation}
\min_{x_{i,t}} \bigg(\frac{1}{2}\alpha_{1,i,t} x_{i,t}^2 +\alpha_{0,i,t}x_{i,t}\bigg) -\lambda_tx_{i,t}.
\end{equation}
From the first order optimality condition, the optimal demand reduced by the DR customer in response to $\lambda_{t}$ is:
\begin{equation}
\label{eq:x_DR}
x_{i,t}^{*} (\lambda)=\beta_{1,i,t}\lambda_{t} + \beta_{0, i,t},
\end{equation}
where coefficients $\beta_{1,i,t} =\frac{1}{\alpha_{1,i,t}}$ and $\beta_{0,i,t} =-\frac{\alpha_{0,i,t}}{\alpha_{1,i,t}}$ are the price sensitivity of DR customer $i$ at DR event $t$.

\subsection{Utility/Aggregator Model}
\label{sec:utility_aggregator_model}
A rational DR customer reduces its power consumption for a given $\lambda_t$ according to the optimality condition in Eq.~\eqref{eq:x_DR} \cite{vladimir}.  The utility/aggregator expects all their DR customers to reduce their power consumption as follows:
\begin{equation}
\label{eq:expectation}
    \mathbb{E}[x_{i,t}(\lambda)]= x_{i,t}^{*}(\lambda)=\beta_{1,i,t}\lambda_t + \beta_{0,i,t}, \quad \forall i=1,\ldots,N,
\end{equation} 
where $\mathbb{E}$ is the expectation operator and $N$ is the number of DR customers participating in the DR program. However, this optimal response is subject to exogenous disturbances pertaining to behavioral aspects of the power consumption of each DR customer and changes in power grid operations (e.g., due to demand forecasting errors). Thus, the utility/aggregator does not  know $x_{i,t}$ exactly, and hence, can use ML techniques to internalize this stochasticity in the DR. The actual demand reduced by a DR customer $i$ during a DR event at time $t$ is:
\begin{equation}
\label{eq:x_DR_opt}
    x_{i,t}(\lambda)= \mathbb{E}[x_{i,t}(\lambda_t)] +\varepsilon_{i,t}, \quad \forall i=1,\ldots,N, 
\end{equation}
where $\varepsilon_{i,t}$ is an exogenous disturbance (error) added to the optimal response provided by the customer $i$ for the DR event at time $t$. The utility/ aggregator observes  $\varepsilon_{i,t}$ if and only if vector ${\boldsymbol{\beta}_{i,t}}=[\beta_{0,i,t}~ \beta_{1,i,t}]^{\top}$ is known to them. However, the utility/aggregator does not know the vector $\boldsymbol{\beta}_{i,t}$ a priori and learn them \cite{robert}. The utility/aggregator must learn the elements of vector ${\boldsymbol{\beta}_{i,t}}$, $\forall i=1,\ldots,N,$ (\textit{exploration}) to design a proper DR incentive $\lambda_t$ at time $t$ (\textit{exploitation}), as shown in Fig. \ref{fig:attack}. \textit{Exploration} and \textit{exploitation} can happen simultaneously \cite{khezeli2017risk,li2017distributed,robert}. We show a way to learn $\boldsymbol{\beta}_{i,t}$ as a function of  $\lambda_t$ and $x_{i,t}$.

\subsubsection{Online Exploration of the Utility Function of DR Customers}
\label{sec:learn_utility_function_DR}
In the set up of Fig.~\ref{fig:attack}, the utility or  aggregator can  sequentially record broadcasted DR incentives and reductions in power consumption  of the DR customers in response to the incentives. Using this recorded  data, $\mathcal{H}_t =\{\lambda_{\tau},$ $x_{i,\tau}\},~\forall i \in \mathcal{N},~\forall \tau \in \mathcal{T}$, where $\mathcal{{N}} =\{1,\ldots, N\}$, $\mathcal{T} =\{1,\ldots, t-1\}$, $|\mathcal{T}|=T$, and $|\mathcal{N}|=N$, the utility or aggregator estimates $\boldsymbol{\beta}_{i,t}$ by minimizing the Mean Squared Error (MSE) of the expected response of DR customers given by Eq.~\eqref{eq:x_DR_opt}. This objective can be parameterized by the empirical loss function: 
\begin{subequations}
\label{eq:MSE}
\begin{align}
   \label{eq:loss}
    L_{i,t}((\lambda,x);\boldsymbol{\beta})= \frac{1}{2T}\sum_{\tau=1}^{T}(x_{i,\tau}-\beta_{1,i,\tau}\lambda_{\tau} -\beta_{0,i,\tau})^2 .
\end{align}
The first order optimality condition of $L((\lambda,x); \boldsymbol{\beta})$ over $\boldsymbol{\beta}_{i,t}$ is expressed as:
\begin{align}
\label{eq:partial_R1}
   \frac{\partial L_{i,t}}{\partial \beta_{1,i,t}} &=\frac{1}{T}\sum_{\tau=1}^{T} (x_{i,\tau}\lambda_{\tau}-\beta_{1,i,\tau}\lambda_{\tau}^2-\beta_{0,i,\tau}\lambda_{\tau}) =0, \\
   \label{eq:partial_R2}
   \frac{\partial L_{i,t}}{\partial \beta_{0, i, t}} &=\frac{1}{T}\sum_{\tau=1}^{T} (x_{i,\tau}-\beta_{1,i,\tau}\lambda_{\tau}-\beta_{0,i,\tau}) =0.
\end{align}
Solving Eqs.~\eqref{eq:partial_R1} and \eqref{eq:partial_R2} and setting $ \frac{1}{T}\sum_{\tau=1}^{T} x_{i,\tau} =\hat{x}_{i,t}$  and $ \frac{1}{T}\sum_{\tau=1}^{T} \lambda_{\tau} =\hat{\lambda}_{t}$, it follows:
\allowdisplaybreaks
\begin{align}
\label{eq:beta1}
    \beta_{1,i,t} &=\frac{\frac{1}{T}\sum_{\tau=1}^{T}x_{i,\tau}\lambda_{\tau} -\hat{x}_{i,t}\hat{\lambda}_{t}}{\frac{1}{T}\sum_{\tau=1}^{T}\lambda^2_{\tau}-\hat{\lambda}^2_{t}},\\
    \label{eq:beta0}
    \beta_{0,i,t} &=\hat {x}_{i,t}-\beta_{1,i,t}\hat{\lambda}_{t}.
\end{align}
Since set $\mathcal{H}_t$  is updated at every DR event at time $t$, Eqs.~\eqref{eq:beta1} and \eqref{eq:beta0}  use \textit{streaming} historical data to estimate $\boldsymbol{\beta}_{i,t}$. This is computationally unattractive for all DR customers over time due to the addition of new data points $\{x_{i,t}, \lambda_t\}$  at each DR event at time $t$. To overcome the computational complexity, the utility or aggregator can use the online gradient descent to update $\boldsymbol{\beta}$ for a DR event at time $t+1$ as:
\begin{align}
 \label{eq:ogd}
     \boldsymbol{\beta}_{i,t+1} = \boldsymbol{\beta}_{i,t}- {\eta_l}(\boldsymbol{\beta}^{\top}_{i,t}\boldsymbol{\lambda}_{t} -x_{i,t})\lambda_{t},
\end{align}
\end{subequations}
where ${\eta_l}$ is a user-defined model learning rate and vector $\boldsymbol{\lambda}_{t} = [1 ~ \lambda_{t}]^{\top}$. In the setting of Eq.~\eqref{eq:ogd}, $\boldsymbol{\beta}$ obtained in Eqs.~\eqref{eq:beta1} and \eqref{eq:beta0} can be realized as $\boldsymbol{\beta}_{i,t}$, which is updated at time $t+1$ to $\boldsymbol{\beta}_{i,t+1}$ based on the new sample set $\{x_{i,t}, \lambda_t\}$ of the DR event at time $t$.

\subsubsection{Online Exploitation of the DR Incentives} 
\label{sec:pricing_design}
This section builds up the DR incentive design mechanism adopted from \cite{li2017distributed} and similar to \cite{robert, khezeli2017risk}. Using the learning model in Section \ref{sec:learn_utility_function_DR}, the utility or aggregator can design DR incentives as shown in Fig.~\ref{fig:attack}.  We set up the \textit{exploitation stage} of the DR learning problem from the aggregator viewpoint (a more general case). The aggregator mediates between DR customers and the utility and offers its capacity to the utility at a pre-defined price ahead of DR event. During the DR event, the aggregator  must supply the offered capacity from its portfolio of DR customers, while  maximizing its DR revenue and minimizing the \textit{dissatisfaction} to the DR customers. For a DR event at time $t$, objective of the aggregator is:
\begin{subequations}
\begin{align}
 \notag
&\min_{x}\overbrace{\sum_{i=1}^N
\frac{1}{2N}\mathbb{E}\bigg[\bigg(\kappa\sum_{i=1}^{N}(x_i+\varepsilon_i) -\kappa D\bigg)^2\bigg]}^{\text{Penalty}}-\overbrace{\frac{\gamma}{N}\mathbb{E} \sum_{i=1}^N(x_i+\varepsilon_i)}^{\text{Revenue}}\\ \label{eq:utility_optimization_function}
&+\underbrace{\frac{1}{N}\mathbb{E}\bigg(\frac{1}{2}\alpha_{1,i}(x_i +\varepsilon_i)^2 +\alpha_{0,i}(x_i+\varepsilon_i)\bigg)}_{\text{Discomfort of DR customers}}, 
\end{align}
where we omitted subscript $t$ as we focus on one particular DR event in rest of the this section. The first  term in Eq.~\eqref{eq:utility_optimization_function} accounts for the penalty of not delivering the committed DR capacity to the utility ($D$~kW), the second term represents the revenue collected by the DR aggregator from the utility for the provided DR services, and the third term is the expected utility function of the DR customers. Notably, these terms are normalized by a factor of $N$ for mathematical tractability. Parameters $\kappa$ and $\gamma$ denote the per-unit penalty price and the per-unit revenue received from providing DR services to the utility, respectively.  Also, disturbance $\varepsilon_{i}$ is independent of other customers and events, and hence,  
is assumed to follow a normal distribution $(\varepsilon_i \sim \mathcal{N}(0,\sigma^2))$, 
i.e., $\mathbb{E}({\varepsilon_{i}})=0$ and variance $Var(\varepsilon_{i})=\sigma_{\varepsilon_i}^2$ \cite{mathieu2011examining,li2017distributed,khezeli2017risk,robert}.

Using the first order optimality condition in Eq.~\eqref{eq:utility_optimization_function}, the optimal demand reduced by a DR customer $i$ is:  
\begin{align}
\label{eq:x_solution}
    x_i^{*} =\beta_{1,i}\bigg[\frac{\kappa D+\gamma-\kappa\sum_{i=1}^N\beta_{0,i}}{1+\kappa\sum_{i=1}^N \beta_{1,i}} +\frac{\beta_{0,i}}{\beta_{1,i}}\bigg],
\end{align}
where $\beta_{1,i} =\frac{1}{\alpha_{1,i}}$ and $\beta_{0,i} =-\frac{\alpha_{0,i}}{\alpha_{1,i}}$.
The DR incentive that the aggregator needs to set for the optimal power reduction in Eq.~\eqref{eq:x_solution} can be determined by replacing $\sum_{i=1}^{N}x_i$ in Eq.~\eqref{eq:utility_optimization_function} by an auxiliary variable $Q$, i.e., $Q= \sum_{i=1}^{N}x_i$. This replacement constructs an equality constraint for the aggregator to satisfy its DR commitment, and the dual variable of this constraint ($\hat \lambda$) returns the optimal DR incentive \cite{li2017distributed}. Thus, Eq.~\eqref{eq:utility_optimization_function} is recast as the following constrained optimization:
\begin{align}
    \notag
&\min_{x,Q}
\frac{1}{2N}\mathbb{E}\bigg[\bigg(\kappa Q+\kappa\sum_{i=1}^{N}\varepsilon_i -\kappa D\bigg)^2\bigg]-\gamma \frac{1}{N}\mathbb{E} \sum_{i=1}^N(x_i+\varepsilon_i)\\ \notag
&+\frac{1}{N}\mathbb{E}\bigg(\frac{1}{2}\alpha_{1,i}(x_i +\varepsilon_i)^2 +\alpha_{0,i}(x_i+\varepsilon_i)\bigg),\\
&\text{s.t.}\quad Q- \sum_{i=1}^{N}x_i =0 : \quad (\hat \lambda).
\label{eq:dual}
\end{align}
 Upon solving Eq.~\eqref{eq:dual}, the optimal response is obtained as:
\begin{align}
\label{eq:x_dual}
    x_i^{*}(\hat \lambda)=\beta_{1,i} \bigg(N\hat \lambda +\gamma + \frac{\beta_{0,i}}{\beta_{1,i}}\bigg).
\end{align}
Comparing Eq.~\eqref{eq:x_solution} with Eq.~\eqref{eq:x_dual}, and indexing for DR event at time $t$,  the optimal value of $\hat \lambda_t$ is given by:
\begin{align}
\label{eq:price}
    \hat \lambda^{*}_t =\bigg[\frac{\kappa D_t-\gamma\kappa \sum_{i=1}^N \beta_{1,i,t} -\kappa\sum_{i=1}^N\beta_{0,i,t}}{N+\kappa N\sum_{i=1}^N \beta_{1,i,t}} \bigg].
\end{align}
\end{subequations}
Since $\hat \lambda^{*}_t$ in Eq.~\eqref{eq:price} is derived from a normalized formulation in Eq.~\eqref{eq:utility_optimization_function}, the actual $\lambda_t$ that needs to be  broadcast to  DR customers is $N\hat \lambda^{*}_t$. The expression in Eq.~\eqref{eq:price} parameterizes the \textit{exploitation stage} of $\lambda^{*}_t$, which can be derived and leveraged by an attacker to launch a causative attack on the DR learning deployed by utility or aggregator. 

\textit{Remark 1:} Eq.~\eqref{eq:price} yields a time-varying value of $\lambda_t$ designed by the utility/aggregator. An attacker can compromise this incentive signal and intentionally deviate DR customers from their optimal power consumption reduction. Further, this  attack can feed erroneous values of $\lambda_t$ and $x_{i,t}$ to a ML algorithm used by the DRAS, which will in turn lead to  sub-optimal values of  $\lambda_t$  for future DR events. However, if the utility/aggregator elects to use fixed (time-invariable) values of $\lambda$ (e.g., for price-insensitive DR customers with constant cost functions) without using a ML algorithm leveraging historical $x_{i,t}$ and $\lambda_t$, there will be no causative attack on the learning process for DR incentives. Additionally, in the case of time-invariable and a priori known values of $\lambda$, the utility/aggregator need not broadcast $\lambda$ in real-time. However, even in this case, some attacks  on DR  signals are still possible (e.g, on DR time and duration-- see PR. 3, 3', c, and c' in Fig.~\ref{fig:DR_communication}). Although time-invariable and a priori known DR incentives may have a smaller attack surface than the case with time-variable incentives considered in this paper, the latter one  dominates across DR programs as it offers greater economic advantage to both the utility/aggregator and DR customers \cite{burger2017review, robert, li2011optimal}.

\section{Attack Design and Implementation}
\label{attack}
\subsection{Devising an Attack}
\label{sec:attack_devise}
 The attacker can launch causative attacks in an online or offline fashion. In the former attack, erroneous data on $\lambda_t$ and $x_{i,t}$ are sequentially fed to the ML algorithm operationalized on the DRAS by launching false data injection attacks on the DR schedules, DR incentives, and customer responses to  DR calls. Due to the price sensitivity of power reductions to these manipulations, as given by a relationship in Eq.~\eqref{eq:x_DR}, the DRAS is caused to involuntarily produce erroneous outcomes on  DR schedules and incentives. In the offline attack, the attacker compromises the DRAS and manipulates the historical training data set on $\lambda_t$ and $x_{i,t}$.  In either case, the ML algorithm used in the DRAS learns an erroneous customer behavior and designs a suboptimal DR incentive.

\subsection{Impact} 
\label{sec:attack_impact_theory}Attacks on DR learning are of interest to both state and non-state actors; however, their goals and motives differ. For example, state actors may employ such attacks to compromise operations of the national power grid, a critical infrastructure system, while  non-state (individual or group) actors can be driven by ransomware. As a result, both attacks from  state and non-state actors can inadvertently or intentionally reduce attractiveness of DR programs. For example, such attacks not only can make the utility/aggregators apathetic to continuing existing and rolling-out new DR programs, but also contribute to  imposing various operational challenges (e.g., frequency and voltage excursions). Furthermore, such attacks may deteriorate credibility of DR programs among DR customers, thus  discouraging them from participation. 

\subsection{Mathematical Formulation} 
\label{sec:attack_design_mathematical_formulation}
In this section, we describe a mathematical model for designing such aforementioned  cyberattacks and analyzes how many DR events and DR customers need to be compromised. We consider the perspective of an attacker and cast its decision-making process as an optimization problem, to determine the number of DR events and DR customers to launch a causative attack on the ML-based ADR employed by the aggregator presented in Section~ \ref{sec:utility_aggregator_model}.  

Consider an attacker that forces the aggregator to learn the anomalous behavior of DR customers as legitimate responses by tampering with the power curtailment of the compromised DR customers  $x_{i,\tau}\mapsto \tilde{x}_{i,\tau}, ~\forall i \in \mathcal{\tilde N},~\forall \tau \in \mathcal{\tilde T}$, where $\mathcal{\tilde{N}} =\{1,\ldots, \tilde N\}$ and $\mathcal{\tilde T} =\{t,~ t+1,\ldots, \tilde T\}$ are the compromised set of DR customers and events, respectively. Historical DR data $\mathcal{H}_t$ is defined for the DR events up to time $t-1$, and the attacks are devised starting from DR event at $t$. Notably, for the offline attack scheme  described in Section~\ref{sec:attack_devise} we enforce  $\mathcal{\tilde T} = \mathcal{T}$. The optimization can be formulated as a bi-level problem, where the top-level considers the perspective of the attacker and the lower-level considers the perspective of the DR incentive design by the aggregator. Since the objective of the lower-level problem is to return an optimal $\lambda_t$ (derived in Eq.~\eqref{eq:price}), we can express the single-level equivalent as a non-linear, non-convex optimization problem:
\allowdisplaybreaks
\begin{subequations}
\label{eq:opt_attack}
\begin{align}
    \label{eq:upper_level_obj}
    &\min_{\tilde {x}_{i,\tau} \in {\mathbb R^{\tilde N\times \tilde T}}}  \norm{ {\sum_{i}^{N}\boldsymbol{\beta}_{i,\tau}} -\boldsymbol{\beta}^{*}}, \quad \forall \tau \in \mathcal{\tilde T}, \\ \label{eq:beta_update_compromised_customer}
   \notag & s.t. \\ 
    &\boldsymbol{\beta}_{i,\tau +1} = \boldsymbol{\beta}_{i,\tau}- {\eta_l} (\boldsymbol{\beta}_{i,\tau}^{\top}\boldsymbol{\lambda}_{\tau}-{\tilde x}_{i,\tau})\boldsymbol{\lambda}_{\tau},\quad \forall i \in {\cal \tilde N},~\forall \tau \in {\cal \tilde T},\\ \label{eq:beta_update_uncompromised_customer}
     &\boldsymbol{\beta}_{i,\tau+1} = \boldsymbol{\beta}_{i,\tau}- {\eta_l} (\boldsymbol{\beta}^{\top}_{i,\tau}\boldsymbol{\lambda}_{\tau}-{ x}_{i,\tau})\boldsymbol{\lambda}_{\tau},\quad \forall i \in {\cal N}_1,~\forall \tau \in {\cal \tilde T},\\ \label{eq:x_cap_comprmised}
& 0\leq \tilde x_{i,\tau}\leq \overline{\tilde x}_{i}, \quad \forall i \in {\cal \tilde N},~\forall \tau \in {\cal \tilde T},\\ \label{eq:x_cap_uncomprmised}
& 0\leq  x_{i,\tau}\leq \overline{x}_{i}, \quad \forall i \in {\cal N}_1,~\forall \tau \in {\cal \tilde T},\\ 
 \label{eq:aggregated D}
& \abs{D_\tau - \sum_{i=1}^{N_1} x_{i,\tau}-\sum_{i=1}^{\tilde N} \tilde x_{i,\tau}} = \delta_{\tau},\quad \forall \tau \in {\cal \tilde T},\\
\label{eq:x_update_comprmised}
& \tilde x_{i,\tau} = \beta_{1,i,\tau}\lambda_{\tau}^{*}+\beta_{0,i,\tau}, \quad \forall i \in {\cal \tilde N},~\forall \tau \in {\cal \tilde T}, \\
\label{eq:x_update_uncomprmised}
& x_{i,\tau} = \beta_{1,i,\tau}\lambda_{\tau}^{*}+\beta_{0,i,\tau}, \quad \forall i \in {\cal  N}_1,~\forall \tau \in {\cal \tilde T},\\\label{eq:lambda_vector}
& [\boldsymbol{\lambda}_{\tau}] = \begin{bmatrix}
           1 \\
           \lambda_{\tau}^{*} \\
         \end{bmatrix},\quad\forall \tau\in{\cal \tilde T}, \\ \label{eq:lower_level}
         & \lambda^{*}_{\tau} =\bigg[\frac{\kappa D_\tau-\gamma\kappa \sum_{i=1}^N \beta_{1,i,\tau} -\kappa\sum_{i=1}^N\beta_{0,i,\tau}}{1+\kappa \sum_{i=1}^N \beta_{1,i,\tau}} \bigg] \geq 0,
\end{align}
\end{subequations}
where $\mathcal{ N}_1 = \mathcal{N}\setminus \mathcal{\tilde N}$, $\overline{\tilde x}$ and $\overline{x}$ are the maximum power curtailment capacities of compromised and uncompromised DR customers, respectively. 
Eqs.~\eqref{eq:upper_level_obj}-\eqref{eq:lambda_vector} are upper-level formulations of the bi-level problem.   Eq.~\eqref{eq:lower_level} is an optimal return of the lower-level problem. Eq.~\eqref{eq:upper_level_obj} defines the objective of the attacker, where it forces the aggregator to learn the value $\boldsymbol{\beta}^{*}$ defined by the attacker. The attacker can conservatively estimate $\boldsymbol{\beta}^{*}$ based on its knowledge of the aggregated $\boldsymbol{\beta}_{t}:=\sum_{i}^N \boldsymbol{\beta}_{i,t}$. To calculate $\boldsymbol{\beta}_{t}$, an attacker can leverage the aggregated realization of Eq.~\eqref{eq:x_DR} expressed as:
\begin{equation}
    \label{eq:expectation_aggregated}
    X^{*}_t(\lambda):= \beta_{1,t}\lambda_t + \beta_{0,t}, 
\end{equation}
where ${\beta}_{1,t} := \sum_{i=1}^N {\beta}_{1,i,t}$ and ${\beta}_{0,t} := \sum_{i=1}^N {\beta}_{0,i,t}$ and $X_t := \sum_{i=1}^N x_{i,t}$
and then use Eq.~\eqref{eq:MSE}. Interestingly, the  aggregated model only requires  knowledge of $X_t$ and $\lambda_t$, which are publicly available (e.g., via smartphone apps of  DR customers and utility documents \cite{ISONE}). Eqs.~\eqref{eq:beta_update_compromised_customer} and \eqref{eq:beta_update_uncompromised_customer} are update rules for $\boldsymbol \beta_{i,t}$ using the online gradient method discussed in Eq.~\eqref{eq:ogd} for the compromised and uncompromised  DR customers, respectively. Similarly, Eqs.~\eqref{eq:x_cap_comprmised} and \eqref{eq:x_cap_uncomprmised} enforce the bounds of customers' DR capacity. Eq.~\eqref{eq:aggregated D} constrains the total curtailment commitment of the DR aggregator within a tolerable amount of aggregated error $\delta_{\tau}$. Eqs.~\eqref{eq:x_update_comprmised} and \eqref{eq:x_update_uncomprmised} are the curtailed power of compromised and uncompromised DR customers, respectively. 
Eq.~\eqref{eq:lambda_vector} is a vector assignment of an optimal DR incentive.

In the offline attack case, the attacker compromises the DRAS  to  access data set  $\mathcal{H}_t$, hyperparameter $\eta_l$, aggregated DR $\mathcal{D}_t$, and power curtailment limits of DR customers $\overline x_i$ that are needed to solve  Eq.~\eqref{eq:opt_attack}. On the other hand, to solve Eq.~\eqref{eq:opt_attack} in the online attack case, the attacker needs to know $\overline x_i$, $\mathcal{D}_t$, and $\boldsymbol \beta_{i,t}$ in advance. Notably, the attacker may use $\mathcal{H}_t$ to estimate  $\boldsymbol \beta_{i,t}$. Solving the optimization in Eq.~\eqref{eq:opt_attack} requires a two-dimensional search  across sets $\cal \tilde N$ and $\cal \tilde T$ with non-linear and non-convex constraints. This requirement can be fulfilled by either compromising a large fraction of DR customers over a few DR events or by compromising a small fraction of DR customers over a larger number of DR events. This trade-off between the compromised number of DR customers and DR events is idiosyncratic to attackers and the learning scheme employed by the aggregator. To prune the search space and to invoke a pragmatic attack scenario, we solve Eq.~\eqref{eq:opt_attack} by fixing either $\cal \tilde N$ or $\cal \tilde T$. This is in line with the inability of the attacker to compromise a larger number of  DR customers for a large number of DR events.

The ease with which  residential DR customers can be compromised depends on  defense mechanisms employed by DR customers, e.g.,  such as using strong WiFi passwords or two-step authentication. It is realistic to assume that an  attacker is constrained, rather than  omnipotent,  i.e., it cannot compromise  all DR customers during all DR events. Therefore, it is important for an attacker to evaluate the DR customers and  events in terms of their contribution towards  learning $\boldsymbol{\beta}_{i,t}$ or designing $\lambda_t$ in Eq.~\eqref{eq:price}. Then the attacker can selectively compromise some DR customers and DR events, as well as to design its attack based on available DR customer and event data. In the following section, we describe an approach based on the Shapley value theorem that can be used by the attacker to  determine the contribution of each DR customer and event data to the attack and, thus, to pick the most valuable DR customers and events for attack design given a particular attack objective.

\subsection{Valuation of the DR Events}
\label{sec:attack_design_valuation_DR_events}
The information available to the attacker from each DR event has a different impact on the planned attack. To rank the effect of the different DR events on learning $\boldsymbol{\beta}_{i,t}$, we use the Shapley value theorem  \cite{shapley1953value,castro2009polynomial, jia2019towards,ghorbani2019data}, which is a  game-theoretical approach to equitably value participation in a coalition. Since $\boldsymbol{\beta}_{i,t}$ of  customer $i$ is learned using historical DR events prior to the DR event at time $t$, contributions of individual historical DR events towards learning $\boldsymbol{\beta}_{i,t}$ can be viewed as a coalition. Accordingly, equitability of the Shapley value in a causative attack in Fig.~\ref{fig:attack} is important along three dimensions: i) total gain of the coalition should be distributed among its players using a gain distribution function (i.e., learning of the utility function), ii) equal contributions to the coalition are ranked equally with zero value for the  non-contributors,  and iii) the sum of the contributions of a player returned by the multiple gain distribution functions is equal  to the contribution of the player returned by the sum of those functions.

\begin{algorithm}
\label{algo:shapley}
\SetAlgoLined
\KwResult{$\phi_\tau, \quad \forall \tau \in \mathcal{T} =\{1,\ldots,t-1$\}}
 $M = \text{Number of Monte-Carlo permutations}, \hat{\phi}_\tau=0, \quad \forall \tau \in \mathcal{T}=\{1,\ldots,t-1\}, |\mathcal{T}| =T, cnt =0$\;
 \While{$cnt<=M$}{
  Randomly generate $\pi \in \Pi (\mathcal{H}_t^i)$ with probability $\frac{1}{T!}$\;
  \For{$\tau \in 1:T$}{
  
 Calculate $\mathcal{P}_\tau^{\pi}$\;
 
 $\beta_{\tau}^{\pi}:=O_{\beta}(\mathcal{P}^{\pi}_\tau\cup \{\tau\}),~\beta_{\setminus \{\tau\}}^{\pi}:= O_{\beta}(\mathcal{P}_\tau^{\pi})$\;
 
  \eIf{$\mathcal{P}_\tau^{\pi}=\oldemptyset$}{$v_\tau =0$\;}{$v_\tau = \overbrace{\frac{L(\mathcal{H}_t^i)}{L(\mathcal{H}_t^i,\beta_\tau^{\pi})}}^{U(\mathcal{P}_\tau^{\pi}\cup \{\tau\})} - \overbrace{\frac{L(\mathcal{H}_t^i)}{L(\mathcal{H}_t^i,\beta_{\setminus\{\tau\}}^{\pi})}}^{U(\mathcal{P}_\tau^{\pi})}$\; 
  }

  }
   $\hat{\phi}_\tau = \hat{\phi}_\tau+v_\tau$\;
 $cnt = cnt+1$;
 }
 $\phi_\tau=\frac{\hat{\phi}_\tau}{M}, \quad  \forall \tau \in \mathcal{T} =\{1,\ldots,t-1$\}
 \caption{Shapley value of DR events of a customer using Monte-Carlo approximation.}
\end{algorithm}

Let $\mathcal{H}_t^i$ be the historical DR data set for a DR customer $i$ defined over $\mathcal T$ with $|\mathcal{T}| =T$ and  $\mathcal{S}\subseteq \mathcal{H}_t^i$. Let $O_\beta$ be an algorithm deployed by the aggregator to learn $\boldsymbol \beta_{i,t}$ (see Section~\ref{sec:learn_utility_function_DR}) and let $U(\mathcal{S})$ be a gain distribution function that returns a value for any given $\mathcal{S}$. We dropped index $i$ in this section as we are dealing with the $i^{th}$ DR customer. As practiced in the ML security literature \cite{ghorbani2019data}, we used a variant of the learning loss function $L(\cdot)$ as the gain distribution function to calculate the learning loss in a data set using algorithm $O_\beta$ (see Eq.~\eqref{eq:loss}). We define $U(\mathcal{\cdot})$ as the ratio of $L(\mathcal{H}_t^i)$ to $L(\mathcal{\cdot})$. In this setting, the Shapley value of a DR event $\tau \in \mathcal{T}$ is:
\begin{equation}
\label{eq:shapley}
\phi_\tau(\mathcal{H}_t^i, O_{\beta}, U) = \sum_{\mathcal{S}\subseteq \mathcal{H}_t^i\setminus \{\tau\}}\frac{U(\mathcal{S}\cup \{\tau\}) -U(\mathcal{S})}{T\binom {T-1}{|S|}}.
\end{equation}
Determining the value of  $\phi_\tau$  with Eq.~\eqref{eq:shapley} requires computing  $U(\mathcal{S})$ and $L(\mathcal{S})$ over $2^{T-1}$ subsets formed by $\mathcal{H}_t^i\setminus \{\tau\}$. Similarly, $U(\mathcal{S}\cup \{\tau\})$ and $L(\mathcal{S}\cup \{\tau\})$ requires calculation over $2^{T-1}$ subsets. In total, the Shapley value of a DR event $\tau \in \mathcal{T}$ DR events require computing $U(\cdot)$ and $L(\cdot)$  over $2^T$ subsets, i.e., the computation of the Shapley value is  exponential in time. Moreover, this calculation becomes less attractive from the computational viewpoint when multiple DR customers are compromised. Hence, to achieve polynomial computation times, we reformulate Eq.~\eqref{eq:shapley} as \cite{castro2009polynomial}:  
\begin{equation}
\label{eq:shapley1}
\phi_\tau(\mathcal{H}_t^i,O_{\beta},U) = \frac{1}{T!}\sum_{\pi \in \Pi(\mathcal{H}_t^i)} [U(P_\tau^{\pi}\cup \{\tau\}) -U(P_\tau^{\pi})],
\end{equation}
 where $\pi$ is a permutation randomly sampled out of all possible $T!$ permutations of DR events $\Pi(\mathcal{H}_t^i)$ and $\mathcal{P}_\tau^{\pi}$ is the set of DR events preceding the DR event at time $\tau$ in $\pi$. This reformulation calculates the marginal contribution of a new DR event at time $\tau$ in learning $\boldsymbol{\beta}_{i,t}$, when added to an existing set of DR events. The new set of DR events can be ordered in $T!$ ways. Averaging the marginal contributions of the DR event at time $\tau$ over these $T!$ ordered sets yields the Shapley value of the DR event. Computing $U(\cdot)$ in Eq.~\eqref{eq:shapley1} over $T!$ ordered sets to approximate $\phi_\tau$ can be avoided by using Monte-Carlo permutations of the order of $T$ \cite{ghorbani2019data, castro2009polynomial}. However, the convergence  of $\phi_\tau$  depends on the distribution of $\mathcal{H}_t$. Algorithm \ref{algo:shapley}  calculates the Shapley value of DR events in polynomial time using the Monte-Carlo approximation~\cite{castro2009polynomial}:
  \begin{itemize}
     \item The Monte-Carlo permutations $M$ are selected so that Shapley values of DR events converge in $M$ iterations.
     \item In each iteration $cnt \in M$, $\pi$ is randomly sampled out of permutations $\Pi(\mathcal{H}^i_t)$. For  a sampled $\pi$, $\mathcal{P}_\tau^{\pi}$ is determined for all DR events. Then $\beta_\tau^\pi$ and $\beta_{\setminus \{\tau\}}^{\pi}$ are calculated for each DR event using the learning algorithm in Section~\ref{sec:learn_utility_function_DR}. If $\tau$ is the first element in $\pi$, $\mathcal{P}_\tau^{\pi} \in \oldemptyset$ for the DR event at time $\tau$, which implies zero Shapley value for the DR event at time $\tau$ in the current iteration $cnt$. If $\mathcal{P}_\tau^{\pi} \not\in 0$, the Shapley value of the DR event is computed as the difference of the ratio of learning losses. $\mathcal{H}_t^i$ is used as the test dataset to assess the learning loss. ${\frac{L(\mathcal{H}_t^i)}{L(\mathcal{H}_t^i,\beta_\tau^{\pi})}}$ and $\frac{L(\mathcal{H}_t^i)}{L(\mathcal{H}_t^i,\beta_{\setminus\{\tau\}}^{\pi})}$ are the ratio of learning losses in the test data set in the presence of the DR event at time $\tau$ and in the absence of the DR event at time $\tau$, respectively.
    \item $M$ Shapley values for each DR event are averaged to obtain the Shapley value of the DR event.
 \end{itemize}

\subsection{Valuation of the DR Customers}
\label{sec:attack_design_valuation_DR_customers}
 At any DR event at time  $t$, the aggregated behavior of DR customers  is defined as: $\boldsymbol{\beta}_t := \sum_{i=1}^N \boldsymbol{\beta}_{i,t}$, i.e., ${\beta}_{0,t} := \sum_{i=1}^N {\beta}_{0,i,t}$ and ${\beta}_{1,t} := \sum_{i=1}^N {\beta}_{1,i,t}$ as explained in Section~\ref{sec:attack_design_mathematical_formulation}.
This implies that the effort of the  aggregator to learn the aggregated behavior of the DR customers is the aggregation of efforts to learn individual DR customer behavior. Thus,  ranking each DR customer  is straightforward, unlike  ranking DR events  in Section~\ref{sec:attack_design_valuation_DR_events}. This is because the gain distribution function $U(\cdot)$ of the aggregated DR customer is a linear addition of the independent behavior of  DR customers, while the function $U(\cdot)$ of the latter involves non-linearities in Eq.~\eqref{eq:MSE} (e.g.,  $\boldsymbol{\beta}_{i,t}$ depend on the preceding DR events). Hence, the contribution of  DR customers can be evaluated without using the Shapley value theorem. Since the behavior of a DR customer is not explicit to either $\beta_{0}$ or $\beta_{1}$, we  consider their combined effect in characterizing the behavior of the customer. Thus, similar to the ranking of DR events, the ratio of learning losses is used to rank DR customers as:
 \begin{equation}
 \label{eq:DR_loss}
     \phi_{i,t} = \frac{L(\mathcal{D}_t;\boldsymbol{\beta}_t)}{L(\mathcal{D}_t;\boldsymbol{\beta}_{i,t})}, \quad \forall i \in 1,...,N,
 \end{equation}
 where $\phi_{i,t}$ is the marginal contribution or the Shapley value equivalent of a DR customer $i$ at DR event at time $t$ in learning $\boldsymbol{\beta}_t$. $\mathcal{D}_t =\{X_t, \lambda_t\}, \quad \forall t \in \mathcal{T}$ is the historical dataset with aggregated power curtailment of DR customers and DR incentives. $L(\mathcal{D}_t;\boldsymbol{\beta}_t)$ and $L(\mathcal{D}_t;\boldsymbol{\beta}_{i,t})$ are the learning losses incurred by the  learned behavior of aggregated DR customers $\boldsymbol{\beta}_t$ and by the learned behavior of $i^{th}$ DR customer $\boldsymbol{\beta}_{i,t}$ on historical data set $\mathcal{D}_t$, respectively. 
 
Equipped with the Shapley value of  DR events and DR customers, the attacker can rank all available data on DR customers and events based on their contribution towards learning  $\boldsymbol{\beta}_{i,t}$. Thus, the attacker may only need to compromise some DR customers and events for the attack to  succeed.
 
 \section{Case Study}
\subsection{Data Collection and Synthesis}
 \label{sec:case_study}
\begin{figure}[!t]
  \centering
\includegraphics[width=\columnwidth, clip=true, trim= 0mm 0mm 0mm 0mm]{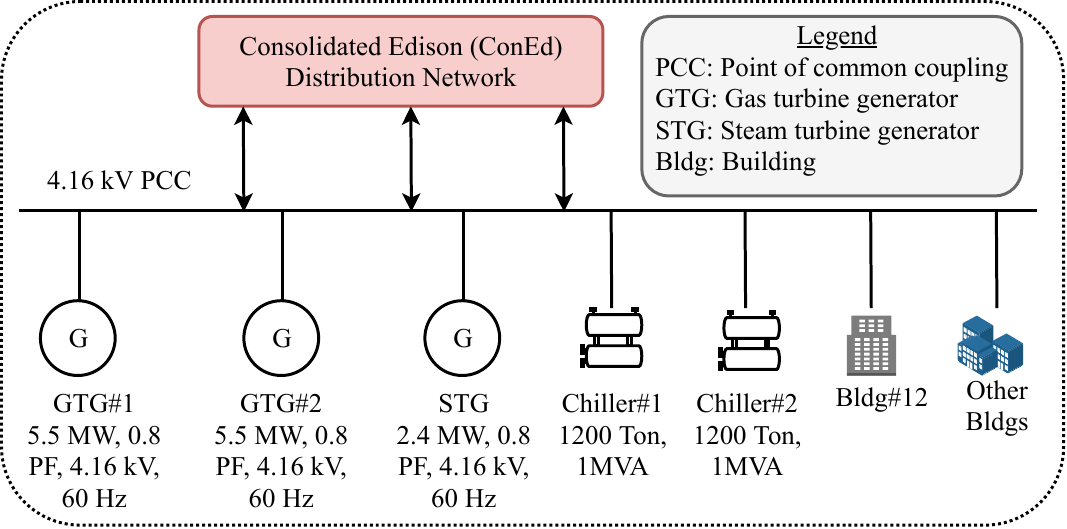}
 \caption{A schematic diagram of the NYU microgrid adopted from \cite{hassan2020hierarchical}.}
 \label{fig:nyu_microgrid_schematic}
 \end{figure}

The case study uses the DR data for building \#12 of New York University (NYU), which is a part of the NYU microgrid shown in Fig.~\ref{fig:nyu_microgrid_schematic}  and is participating in the DR program organized by Consolidated Edison, the local electric power utility in NY~\cite{hassan2020hierarchical, smartAC_conED}. The aggregated DR of this building with over 20 DR events was randomized for 50 individual DR customers such that the power reduction of each DR customer is between $5$ kW-$50$ kW and the DR incentive is between $1$ \$/kWh-2 \$/kWh. The capacity and incentives observed during these 20 DR events are given as $\mathcal{H} =\{\lambda_{t}$, $x_{i,t}, ~i=1, \ldots, 50, t=1, \ldots, 20$\}. The case study is carried out on a MacBook Air with a 2.2 GHz Intel Core i7 processor and 8 GB RAM. The  non-convex and non-linear optimization in Eq.~\eqref{eq:opt_attack} is solved in Julia using Ipopt. All simulations instances described below were solved within  one hour, which is sufficient to accommodate the time frame for  DR event planning (e.g., Consolidated Edison sends their public, day-ahead advisory notice to DR customers 21 hours or more prior to  a call window \cite{coned_2020}).

\subsection{Valuation of Historical Data}
\label{sec:val_results_historical_data}
\begin{figure}[!t]
  \centering
  \vspace{-3mm}
\includegraphics[width=\columnwidth, clip=true, trim= 4mm 0mm 8mm 5mm]{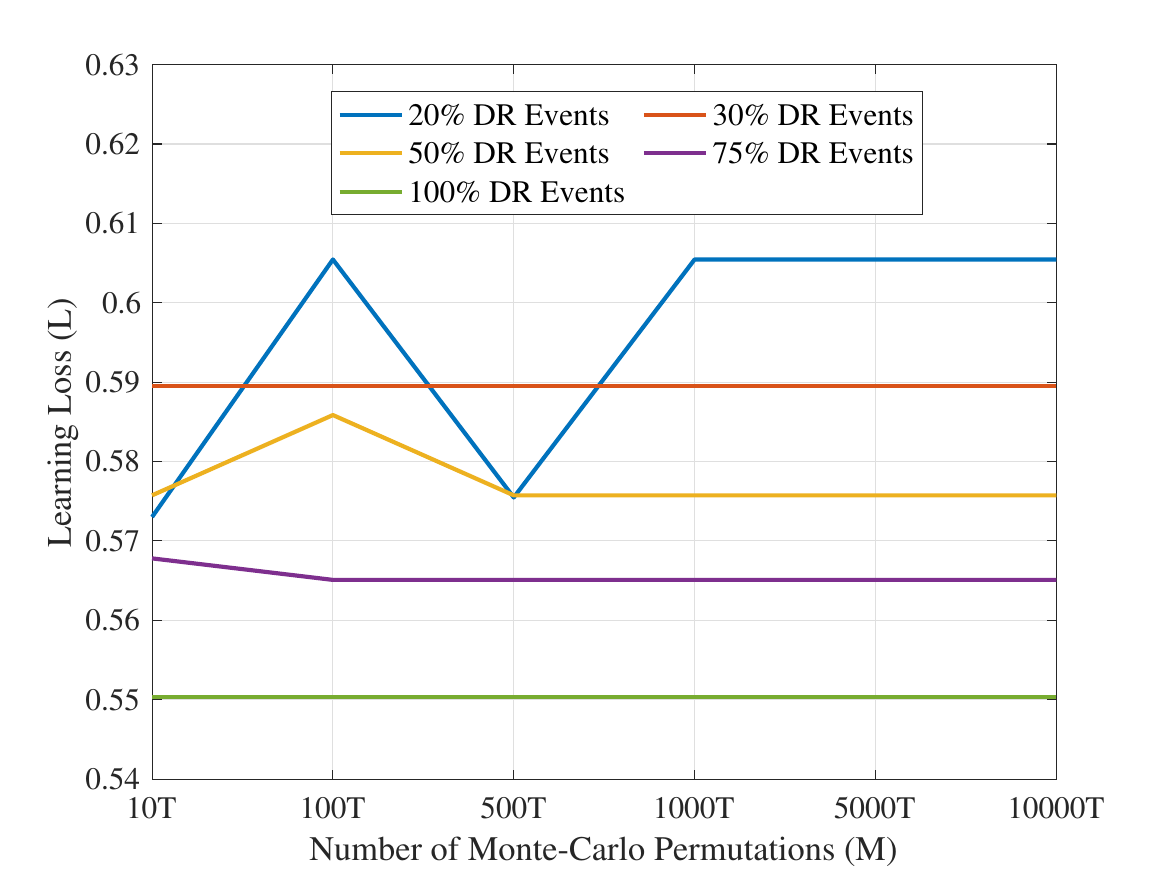}
 \caption{Convergence of the learning loss of a randomly selected DR customer for $(\cdot)$\% DR events with the  greatest relative Shapley value over permutations. $T$ is the number of DR events.}
 \label{fig:shapley_learning_loss}
 \end{figure}

\begin{figure}[!t]
  \centering
\includegraphics[width=1\columnwidth, clip=true, trim= 7mm 0mm 10mm 7.5mm]{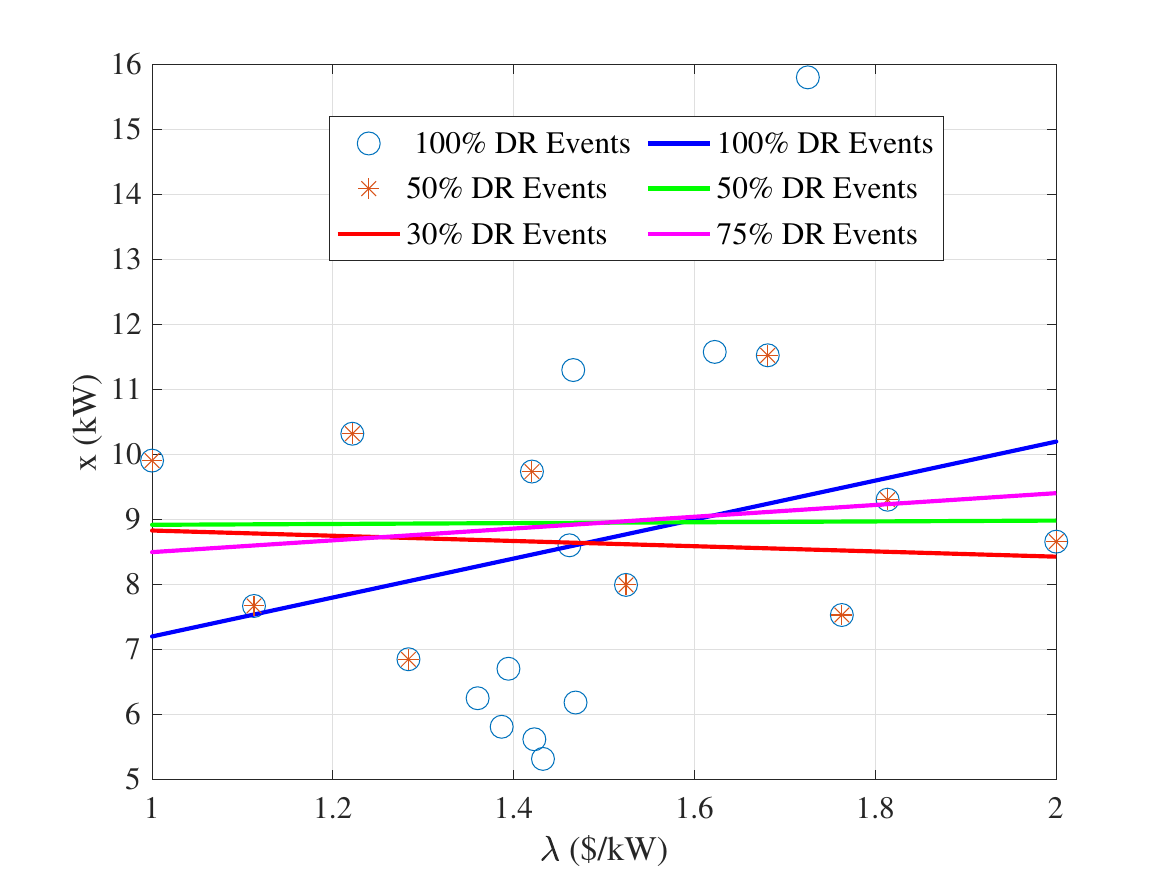}
 \caption{Convergence of the learned behavior of a randomly selected DR customer for $(\cdot)$\% DR events with the greatest relative Shapley value. The circles and stars represent the DR events and the lines display the behavior of a randomly selected DR customer. The same DR customer and its DR events are used as in Fig.~\ref{fig:shapley_learning_loss}.}
 \label{fig:shapley_regressor}  
 \end{figure}

In this section, we investigate the effect of the number of historical DR events and DR customers on learning their behavior. 
First, we evaluate  contributions of individual historical DR events towards learning the behavior of DR customers. Fig.~\ref{fig:shapley_learning_loss} presents the MSE learning losses for the DR events of a randomly selected DR customer $i$ over randomly generated Monte-Carlo permutations. The learning loss in Fig.~\ref{fig:shapley_learning_loss} is achieved in two steps. First, the contribution of each historical DR event in learning the behavior of each DR customer is ranked using the Shapely value theorem from Section ~\ref{sec:attack_design_valuation_DR_events}. That is, the greater the Shapley value of that DR event, the greater  contribution  it makes towards learning the actual value of $\boldsymbol{\beta}_{i,t}$. Second, all DR events are sorted in the order of their contribution and $(\cdot)\%$ of the highest Shapley-valued DR events is used to compute the relative MSE learning loss. Since the calculation of Shapley value is a Monte-Carlo-based approximation, the learning losses of the DR events stabilize after $1000 T$ Monte-Carlo permutations as shown in Fig.~\ref{fig:shapley_learning_loss}, where $T$ is the number of time instances with historical DR events. As Fig.~\ref{fig:shapley_learning_loss} shows, the case when $\approx 50\%$ of the most Shapley-valued DR events is available for learning yields learning losses close to the case with $100\%$ availability of DR events. This means that the attacker can  learn aggregated  behavior of DR customers  $\boldsymbol{\beta}_{i,t}$ sufficiently well with only $\approx 50\%$ of DR events. Furthermore, Fig.~\ref{fig:shapley_regressor} shows that the number of DR events available for the attack design does not  significantly change the learned value of $\boldsymbol{\beta}$. Rather the contribution of the DR events in learning the behavior is significant. As a result, the Shapley value tends to be greater for DR events in the proximity of the curve learned from $100\%$ DR events. For example, this can be  seen in Fig.~\ref{fig:shapley_regressor}, where  50\% of the DR events with the greatest  Shapley values are near the curve learned from $100\%$ DR events. 

\begin{figure}[!t]
   \centering
\includegraphics[width=1\columnwidth, clip=true, trim= 7mm 0mm 15mm 10mm]{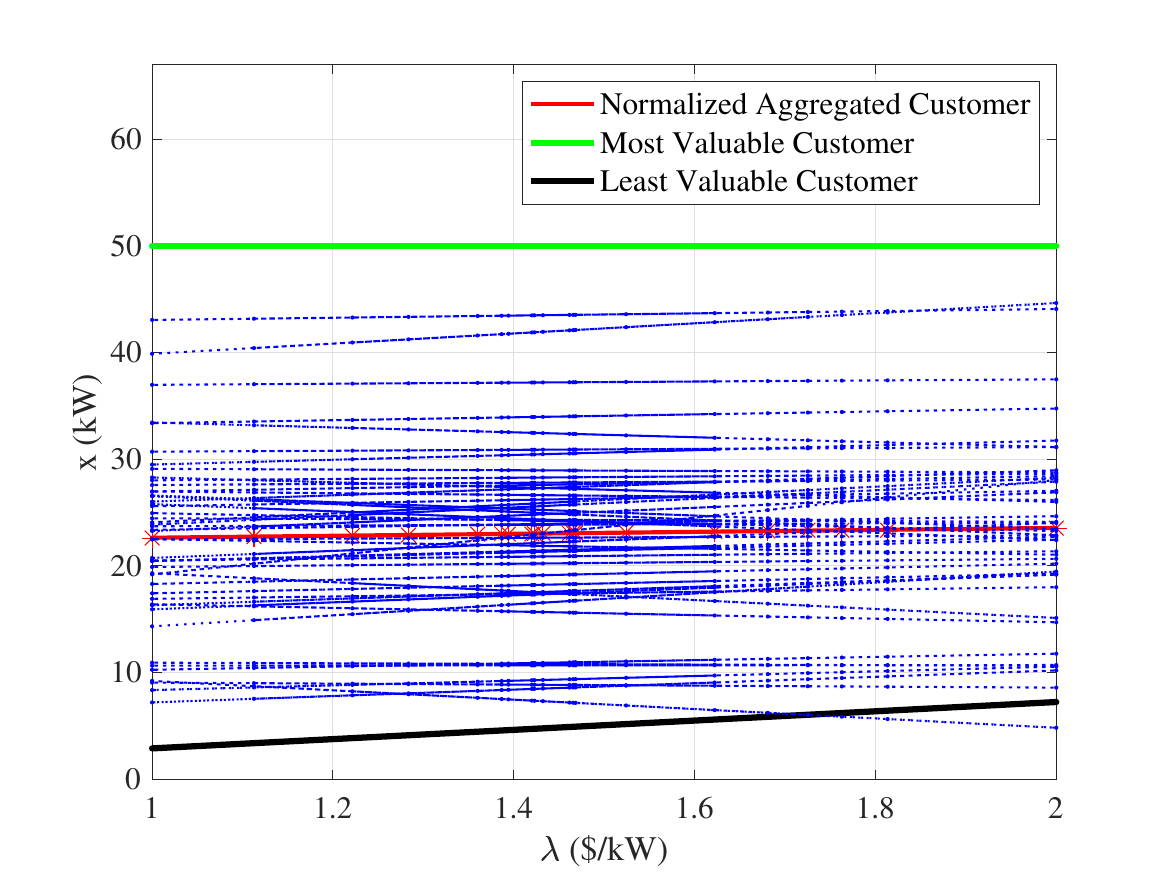}
 \caption{Power curtailment behaviors of the DR customers learned at the DR event at time $t$. The behavior of the dataset $\{X_t\} \in \mathcal{D}_t$ is normalized by the number of DR customers $N$ to visualize all behaviors in the same plot.}
 \label{fig:val_particpants_result}
  \end{figure}

Next, we evaluate individual contributions of DR customers on learning their aggregated behavior. Fig.~\ref{fig:val_particpants_result} presents the learned behavior of all DR customers $\boldsymbol{\beta}_{i,t}$ and their aggregated behavior $\boldsymbol{\beta}_{t}$ at DR event at time $t$. The effect of each DR customer on the aggregated behavior is evaluated using Eq.~\eqref{eq:DR_loss}. The aggregated behavior of the DR customers is normalized by $N$ to visualize their behaviors comparatively. The most valuable DR customer for attack design can be identified by the slope of its behavior (given by $\beta_{1}$), which aligns with the direction of the aggregated behavior. Conversely, the least valuable DR customer has its slope in the opposite direction. The intercept (given by $\beta_{0}$) of the most valuable DR customer is the largest among all DR customers and the least valuable DR customer has the smallest intercept. That is, the most and the least valuable customer has the greatest and lowest values of $\phi_{i,t}$ in Eq.~\eqref{eq:DR_loss}. Therefore, in both offline and online attack cases, the attacker can rank  compromised DR events and  DR customers based on how critical they are for achieving a given objective of the planned attack and how much data is needed for an attack to succeed.

This relationship between DR customers/events data and their contribution towards the attack is established in a data-driven manner, rather than based on  physical attributes (e.g., a type of DR appliances or an underlying energy conversion technology). As a result of the data-driven relationship, the valuation of DR customers/events is case-specific. Thus, drawing a physical intuition behind the physical attributes of the DR appliance and their Shapley value is not straightforward.

 \subsection{Attack Assessment}

\subsubsection{Monetary Attack} \label{sec:monetary_attack} 
\begin{figure}[!b]
   \centering
\includegraphics[width=1\columnwidth, clip=true, trim= 7mm 0mm 15mm 10mm]{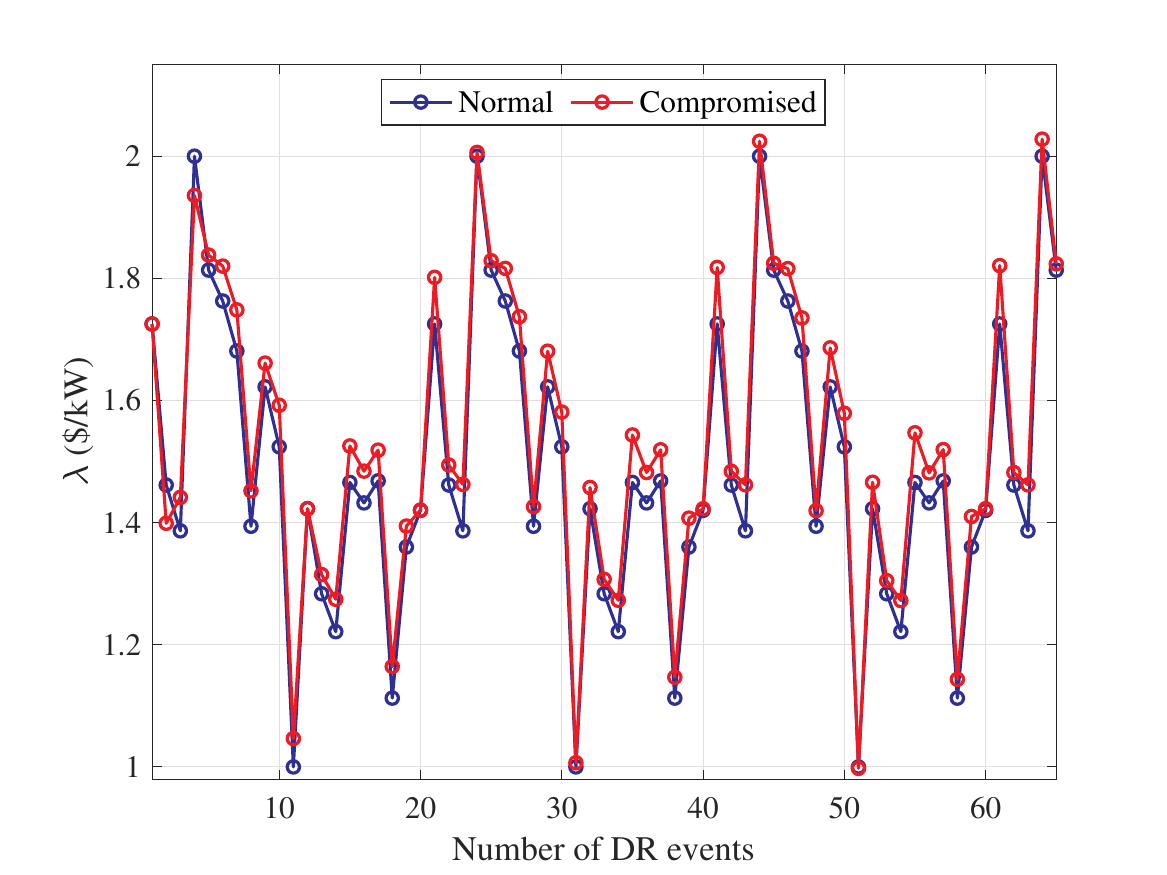}
 \caption{DR incentive $\lambda_t$ broadcast by the utility/aggregator to DR customers during normal and compromised DR events. Note that the aggregated power curtailment delivered  to the utility/aggregator  is the same in the normal and compromised DR cases.}
 \label{fig:dynamic_price_increase}
 \end{figure}

This  section uses the most valued DR customers  and  DR events, obtained as described in Section~ \ref{sec:val_results_historical_data},  to launch a causative attack on the DR learning scheme described in Section~\ref{sec:attack_design_mathematical_formulation}. Consider an attack that manipulates power curtailments of DR customers $x_{i,t}$ to  increase  DR incentives $\lambda_t$  broadcasted for  future DR events, thus making the DR program more costly for the utility/aggregator. To do so, the attacker sets  $\boldsymbol{\beta}^*=[\beta_{0}^{*} ~ \beta_{1}^{*}] ^{\top} =[\beta_{0,t}~ 0.95 \beta_{1,t}]^{\top}$ as its objective in the optimization problem in Eq.~\eqref{eq:opt_attack}. The decrease in $\beta_{1}$ is associated with a DR service that is more expensive to the aggregator and an increase in the incentive to the DR customers. Therefore, this attack can be launched by a customer or a group of customers (see Section~\ref{sec:attack_impact_theory}) to deliberately increase their DR incentives and, thus, payoffs. This increase in the DR incentive  may not be economical for the aggregator/utility and, hence, may lead to a cancellation of the DR call. The attacker remains stealthy by delivering the committed DR capacity to the utility while manipulating the DR customers. 
 At least $\#65$ DR events are needed before an aggregator learns the anomalous behavior of the DR customers (i.e., the target of $\boldsymbol{\beta}^*$ is achieved within  $65$ DR events). About $30\%$ of the DR customers with large Shapely values can achieve the attack objective in the $65$ DR events. 
 Fig.~\ref{fig:dynamic_price_increase} shows the DR incentives $\lambda_t$  during normal and compromised states of DR customers. A considerable number of DR events is necessary for this attack to succeed. However, instead of acquiring a large number of DR events for attack planning, the attacker can operationalize an attack with fewer DR event observations, if the attacker does not seek to  match the power curtailment and DR target amount as  enforced in  Eq.~\eqref{eq:aggregated D}. However, this mismatch observed in real-time by the aggregator/utility may inform them of the attack. Thus, the number of DR events required for attack success  leads to a trade-off with attack stealthiness.
 
\begin{figure}[!t]
   \centering
\includegraphics[width=1\columnwidth, clip=true, trim= 7mm 0mm 15mm 7mm]{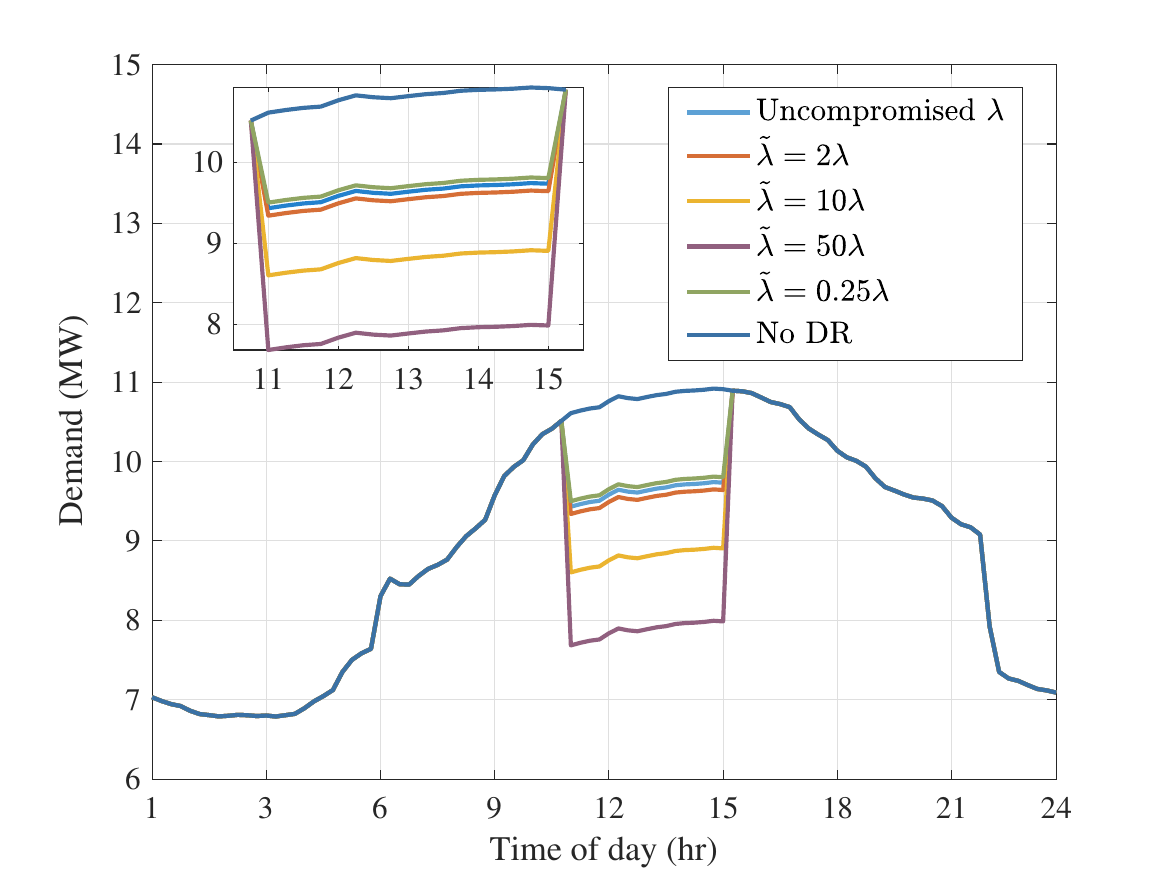}
\caption{Effects of manipulating  $\lambda_t$ on the demand profile in the NYU microgrid.}
 \label{fig:attack_demand_profile_nyu}
\end{figure}
 
\subsubsection{Technical Attack}
As opposed to the monetary attack described above, we shift in this section to technical attacks, which  aim to instantly disrupt electricity supply. To illustrate this attack, we use the  NYU microgrid described in Section~\ref{sec:case_study}. The NYU microgrid has a total generation capacity of 13.4 MW and is capable of islanding from the local distribution network operated by  Consolidated Edison as described in \cite{hassan2020hierarchical}. We consider  a DR event between 11:00 to 15:00 hours. The attacker manipulates a single value of $\lambda_t$ broadcast by the utility/aggregator for that DR event by exploiting the devices and communication channel as shown in Fig.~\ref{fig:attack}. Then, DR customers respond optimally to the compromised DR incentive $\tilde \lambda_t$ as in Eq.~\eqref{eq:x_DR}. In turn, these responses of DR customers alter the  demand of the  microgrid during the DR event window. 
Fig.~\ref{fig:attack_demand_profile_nyu} demonstrates how this change in $\lambda_t$ leads to changes in power curtailments of DR customers. If  the attacker decreases $\lambda_t$ by 4$\times$, the power curtailment is diminished by $\approx$ 3.8\%. On the other hand, if the attacker increases $\lambda_t$ by 50$\times$, the power curtailment of DR customers will increase by $\approx$ 21.5\%. Both an unexpected increase and  decrease in the demand of the microgrid leads to an under- or over-frequency event.

\begin{figure}[!b]
  \centering
  \vspace{-3mm}
\includegraphics[width=\columnwidth, clip=true, trim= 0mm 55mm 0mm 46mm]{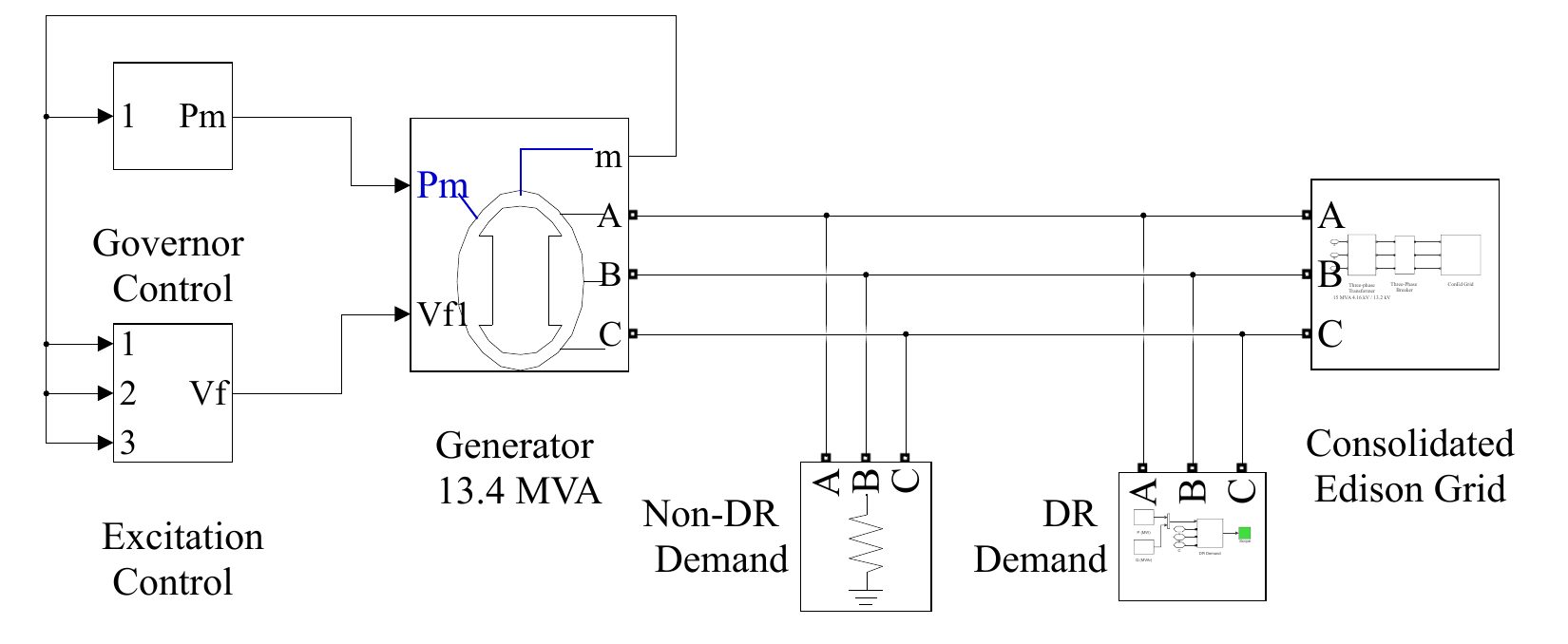}
 \caption{An equivalent MATLAB/Simulink model to simulate performance of the NYU microgrid.}.
 
 \label{fig:nyu_microgrid_matlab}
 \end{figure}
 
\begin{figure}[!t]
  \centering
  \vspace{-3mm}
\includegraphics[width=\columnwidth, clip=true, trim= 5mm 0mm 12mm 0mm]{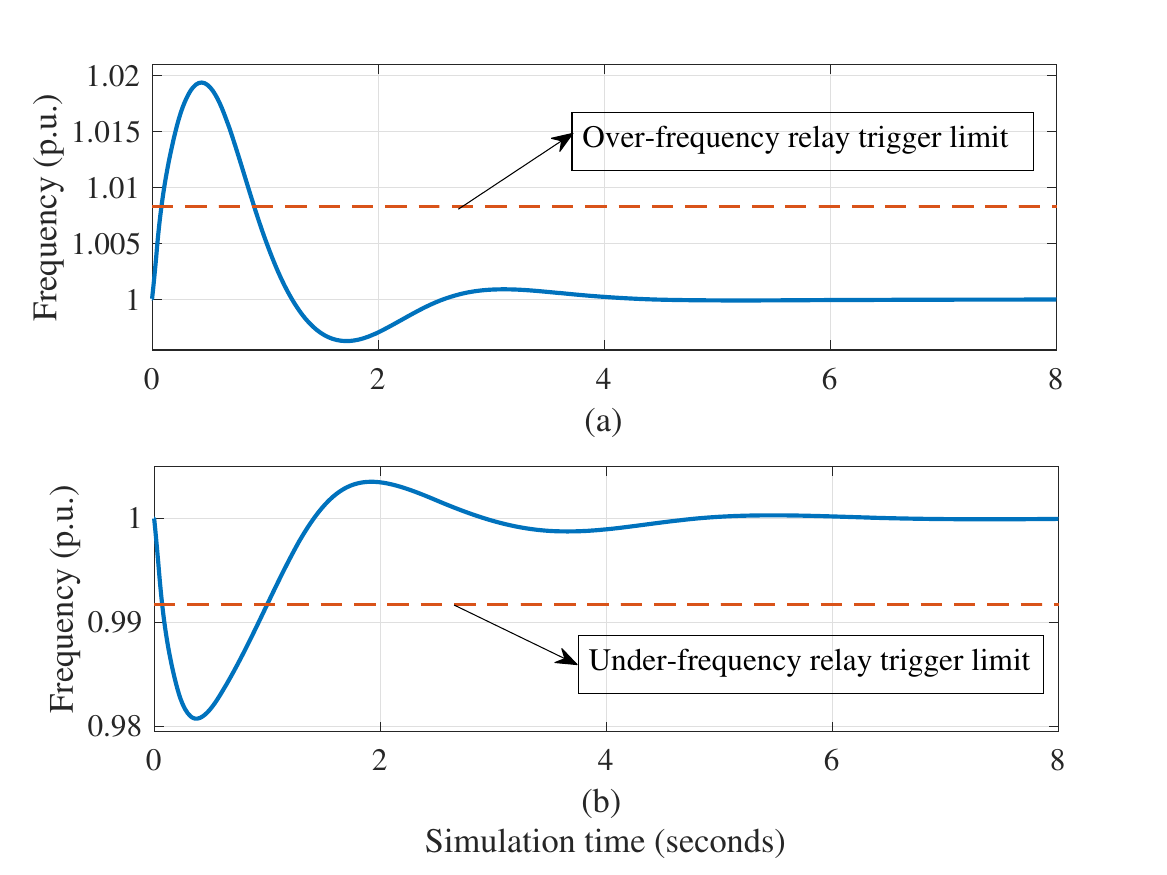}
 \caption{A simulated frequency profile in the NYU microgrid following the compromised DR event with $\tilde \lambda_t = 50\lambda_t$ in Fig. 8: (a) Over-frequency event at the DR start time - 11:00 hours, (b) Under-frequency event at the DR end time - 15:00 hours.}
 \label{fig:nyu_microgrid_freq}
 \end{figure}

To demonstrate the frequency excursions in the NYU microgrid caused by the attacks, the NYU microgrid is modeled in MATLAB/Simulink as shown in Fig.~\ref{fig:nyu_microgrid_matlab}. To make this model tractable, we aggregate three generators into one using an equivalent generator model with a speed governor controller and an excitation controller. The demand on the microgrid is divided into the DR and non-DR demands in these simulations. Since the attack with $\tilde \lambda_t =50\lambda_t$ in Fig.~\ref{fig:attack_demand_profile_nyu} leads to the largest power curtailment (e.g.,  to 7.68 MW at 11:00 hours), we use it in the analysis. If the attack remains undetected, it incurs a sudden and significant demand increase at the DR end time (e.g., to 10.8 MW at 15:00 hours). These demand excursions, in turn, cause an over-frequency event of $\approx$ 1.018 per unit (p.u.) and an under-frequency event of $\approx$ 0.981 p.u. as shown in Figs.~\ref{fig:nyu_microgrid_freq} (a,b) respectively. Since the attack creates sudden substantial changes in the microgrid demand at 11:00 and 15:00 hours, we simulate these two instances to observe the maximum possible frequency excursion in the microgrid.
These frequency excursions can trigger the frequency relays, where the limits are set to 0.9916 p.u.  and 1.0083 p.u. for under- and events over-frequency as recommended by IEEE Standard 1547 \cite{6818982}. Depending on  the range, in which the settings of the under- and over frequency relays can be adjusted, these frequency excursions may island the microgrid from the Consolidated Edison network and trip the microgrid generators. Such excursions can also cause wear and tear on the microgrid components and may propagate into the  systems.
 
\section{Defending the Attack}
The success of the causative attack mechanism described in this paper hinges on the ability of the attacker to tamper with  data on  DR incentives and power curtailments. This data can be infiltrated at various devices and communication channels between DR customers and the utility/aggregator. The use of industry-grade security techniques on DRAS servers and communication channels, e.g., OpenADR, BACnet, CTA-2045, and increased  awareness of cyber hygiene among DR customers can reduce the likelihood of intrusion. In contrast,  idiosyncratic cyber hygiene of DR customers and various possibilities of zero-day vulnerabilities provide additional entry points for attackers. Therefore, defense schemes against such causative attacks are needed to secure ML applications for DR programs. There are several such defense schemes in recent ML security literature \cite{steinhardt2017certified, zhang2017game}. Steinhardt et al. \cite{steinhardt2017certified}  proposed a defense mechanism that aims to remove the compromised data from the original dataset. This data sanitizing process is achieved by estimating the probability distribution of the uncompromised dataset in the attacked dataset. Similarly, Zhang and Zhu \cite{zhang2017game} presented a game-theoretic defense scheme based on a distributed support vector machine, where the defender does not update the algorithm if the compromised data samples are detected. However,  performance of these security mechanisms varies when applied to   a specific engineering  domains and to the best of the authors' knowledge there is no such mechanism designed for DR programs.

\section{Conclusion}
This paper presents a novel attack on the ADR learning used by an utility/aggregator by either providing an erroneous response from the DR customers or by broadcasting anomalous incentives to the DR customers. This causative attack is launched on the ADR framework of an utility/aggregator by learning the quadratic cost function of the DR customers and broadcasting an optimal DR incentive to the customers. As shown, the attack can be launched on the data synthesized from a DR-enrolled building. Such causative attacks on the learning schemes used by utilities/aggregators are realistic especially when there are a larger number of DR events.

\bibliographystyle{IEEEtran}
\bibliography{ref_learning}
\end{document}